\documentclass[hyper]{JHEP} 

\usepackage{epsfig}

\newcommand\fverb{\setbox\pippobox=\hbox\bgroup\verb}

\newcommand\fverbdo{\egroup\medskip\noindent%

            \fbox{\unhbox\pippobox}\ }

\newcommand\fverbit{\egroup\item[\fbox{\unhbox\pippobox}]}

\newbox\pippobox

\title{Remark About Non-Relativistic String in Newton-Cartan Background  and Null Reduction}
\author{J. Kluso\v{n}\\
Department of
Theoretical Physics and Astrophysics\\
Faculty of Science, Masaryk University\\
Kotl\'{a}\v{r}sk\'{a} 2, 611 37, Brno\\
Czech Republic\\
E-mail: \email{klu@physics.muni.cz}} \preprint{}

 \abstract{We analyze non-relativistic string in Newton-Cartan
 background that was found recently in
[arXiv:1705.03535 [hep-th]]. We find its Hamiltonian formulation
and study structure of constraints. We also discuss
a relation between string in Newton-Cartan Background and T-duality along null
reduction.}

\def\hv{\hat{v}}
\def\he{\hat{e}}

\def\ttau{\tilde{\tau}}

\def\tF{\tilde{F}}

\def\tT{\tilde{T}}

\def\teta{\tilde{\eta}}

\def\be{\begin{equation}}

\def\ee{\end{equation}}

\def\bea{\begin{eqnarray}}

\def\eea{\end{eqnarray}}

\def\mH{\mathcal{H}}

\newcommand{\hh}{\hat{h}}

\newcommand{\mG}{\mathcal{G}}

\newcommand{\bT}{\mathbf{T}}

\newcommand{\mL}{\mathcal{L}}

\def\pb #1{\left\{#1\right\}}

\begin{document}
\section{Introduction}
Today it is well known that non-relativistic holography is a  very
useful for description of strongly coupled condensed matter systems,
see for example \cite{Hartnoll:2016apf}. In fact, non-relativistic
holography is an example of non-Lorentzian systems that have been
studying  last few years very intensively. For example, effective
actions of non-relativistic field theories were analyzed in
\cite{Son:2013rqa,Christensen:2013lma,Geracie:2014nka,Jensen:2014aia,Hartong:2014pma}. Further,
non-relativistic local symmetries are crucial for the proposal of
renormalizable theory of gravity known today as Ho\v{r}ava-Lifshitz
gravity \cite{Horava:2009uw}. In fact, recently close relation
between Ho\v{r}ava-Lifshitz gravity and Newton-Cartan gravity was
found  in \cite{Hartong:2015zia}. Finally, three dimensional
non-relativistic gravities were also very intensively studied in
\cite{Bergshoeff:2016lwr,Hartong:2016yrf}.

Certainly it is  very interesting question to analyze extended
object in the context of non-relativistic gravity. Non-relativistic
strings were firstly introduced from different point of view
in \cite{Gomis:2000bd,Danielsson:2000gi}, for further analysis in the context of
string theory, see for example \cite{Kluson:2006xi,Gomis:2005bj,Gomis:2005pg,Gomis:2004pw,Brugues:2004an}.
Non-relativistic strings and p-branes gained renewed interest recently
when these objects were studied in the context of Newton-Cartan gravity and
Carroll gravity
\cite{Bergshoeff:2015wma,Bergshoeff:2014gja,Bergshoeff:2014jla,Andringa:2012uz,
Cardona:2016ytk,Batlle:2016iel,Gomis:2016zur,Kluson:2017fam,Kluson:2017vwp,Harmark:2017rpg,Kluson:2017ufb,Kluson:2017djw,Kluson:2017pzr,Barducci:2017mse,Kluson:2017abm,Kluson:2018uss}.

There are basically two ways how to derive non-relativistic string in Newton-Cartan background. The first one is based on the gauging procedure of
non-relativistic string in the flat background which was introduced in
\cite{Andringa:2012uz}. The characteristic property of this approach is that
 the number of longitudinal dimensions is doubled with respect to standard
Newton-Cartan gravity. This fact also naturally emerges  when we
construct non-relativistic strings or p-branes in Newton-Cartan
background implementing the limiting procedure
\cite{Bergshoeff:2015uaa,Bergshoeff:2015sic}. However there is an
alternative procedure how to define Newton-Cartan theory which is
based upon null dimensional reduction
\cite{Duval:1984cj,Duval:1990hj,Julia:1994bs} of higher dimensional
theory. Then one can ask the question whether null dimensional
reduction of the Polyakov action leads to new action for the string
in Newton-Cartan background. This question was answered in a very
nice paper \cite{Harmark:2017rpg} where  new covariant action for
string in Newton-Cartan background was found. It was further   shown
there that performing the second non-relativistic limit that affects
both target space and world-sheet coordinates leads to  sigma models
that describe strings moving in novel non-Lorentzian geometry. The
main difference between this approach and the construction of
non-relativistic string performed in \cite{Andringa:2012uz} is that
we obtain string moving in Newton-Cartan gravity without doubling
the number of longitudinal directions. This is a  very attractive
property of this construction. However the price that we have to pay
for it is that now there is an additional mode that propagates on
the world-volume of the non-relativistic string. The presence of
this mode is a reflection of the fact that non-relativistic string
in Newton-Cartan background is constructed through
 dimensional reduction from higher dimensional space-time with null isometry.

Since this proposal is very interesting it deserves further study.
The goal of this paper is precisely focused on this analysis. In the
first part we determine Hamiltonian form of the action introduced in
\cite{Harmark:2017rpg}. We find this Hamiltonian as the sum of two
constraints which are manifestly invariant under Galilean
transformations. We further show that these constraints are the
first class constraints with agreement with the fact that the string
action is invariant under world-sheet diffeomorphism.

At the second part of this paper we focus on alternative derivation of the non-relativistic string in Newton-Cartan background that reflects  its
deep string theory origin. In more details, we start with the Hamiltonian formulation of the string  in the background with null isometry. Then we show that when we study string in this background the correct way how to perform dimensional reduction
in case of the fundamental string is to perform T-duality along this direction.
It is well known that  string T-duality
can be interpreted as  canonical transformation
\cite{Alvarez:1994wj,Alvarez:1994dn}. Performing this canonical transformation for the string in the background with null isometry we find that the Hamiltonian constraint takes exactly the same form as the Hamiltonian constraint for the
string in Newton-Cartan background that we found in the first section. Finally we perform an inverse Legendre transformation and we find that resulting Lagrangian density exactly coincides with the Lagrangian found in \cite{Harmark:2017rpg} which is again nice consistency check.

This paper is organized as follows. In the next section (\ref{second}) we review the construction of non-relativistic string in Newton-Cartan background as was performed in \cite{Harmark:2017rpg}. Then in section (\ref{third}) we perform canonical analysis of this theory, determine constraint structure and calculate Poisson algebra of constraints. We also discuss second non-relativistic limit
in the context of Hamiltonian formulation. We also briefly discuss the
gauge fixed theory. In section (\ref{fourth}) we show that non-relativistic string in Newton-Cartan background can be defined starting with the string in the background with null isometry and then performing T-duality along null direction.
Finally in conclusion (\ref{fifth}) we outline our result and suggest possible
extension of this work.

\section{Review of Non-Relativistic String in Newton-Cartan Background}
\label{second} In this section we review the construction of
non-relativistic string in Newton-Cartan background as was presented
\cite{Harmark:2017rpg}. The starting point is the Polyakov action in
general background
\begin{equation}
S=\int d^2\sigma \mL=-\frac{T}{2}\int d^2\sigma
\sqrt{-\gamma}\gamma^{\alpha\beta}g_{\alpha\beta} \ , \quad
g_{\alpha\beta}=\partial_\alpha x^M\partial_\beta x^N G_{MN}  \ ,
\end{equation}
where $G_{MN}$ is $d+2$ dimensional target space time metric, $x^M,M,N=0,
\dots,d+1$ determine position of the string in the target space-time and $T$ is the string tension. Finally, $\gamma_{\alpha\beta}$ is two dimensional world-sheet metric where $\gamma=\det \gamma_{\alpha\beta} \ ,
\gamma_{\alpha\beta}\gamma^{\beta\delta}=\delta_\alpha^\delta$, where $\alpha=0,1$ and where we label world-sheet metric with coordinates $\sigma^\alpha$ so that $\partial_\alpha\equiv \frac{\partial}{\partial \sigma^\alpha}$.

As the next step we consider the target space metric in the form
\begin{equation}
ds^2=G_{MN}dx^M dx^N=2\tau(du-m)+h_{\mu\nu}dx^\mu dx^\nu \ ,
\end{equation}
where $\mu,\nu=0,1,\dots,d, M=(u,\mu)$ and where
\begin{equation}
\tau=\tau_\mu dx^\mu \ , \quad m=m_\mu dx^\mu \ ,
\end{equation}
where $\det h_{\mu\nu}=0$.
The tensors $\tau_\mu,m_\mu$ and $h_{\mu\nu}$ are independent of $u$.
We also define $e_\mu^{ \  a}$ through the relation
\begin{equation}
h_{\mu\nu}=e_\mu^{ \ a}e_\nu^{ \ b}\delta_{ab} \ , \quad  a=1,\dots,d \ .
\end{equation}

Now we are ready to proceed to the analysis
 introduced in \cite{Harmark:2017rpg}.
The main idea is
to remove the field $x^u$ from the description. First of all we define momentum current conjugate to $u$ as
\begin{equation}
P^\alpha_u=\frac{\partial \mL}{\partial (\partial_\alpha x^u)}
=-T \sqrt{-\gamma}\gamma^{\alpha\beta}G_{u\mu}\partial_\beta  x^\mu=
-T \sqrt{-\gamma}\gamma^{\alpha\beta}\tau_\beta \ ,
\end{equation}
where $\tau_\alpha=\tau_\mu \partial_\alpha x^\mu $. Note that the equation of motion for $x^u$ has the form
\begin{equation}
\partial_\alpha \left(\frac{\partial \mL}{\partial_\alpha x^u}\right)=
\partial_\alpha P^\alpha_u=0
\end{equation}
so that the condition of imposing $x^u$ on-shell is equivalent to the condition
\begin{equation}\label{eqPu}
\partial_\alpha P^\alpha_u=0 \ .
\end{equation}
To proceed further  we require that $P^\alpha_u$ is an independent variable which can be imposed by the Legendre transformations
\begin{equation}
\hat{\mL}=\mL-P_u^\alpha \partial_\alpha x^u \ ,
\end{equation}
where $\hat{\mL}$ is independent on $x^u$. On the other hand imposing $P_u^\alpha$ to be on shell implies relation between metric components $\gamma^{\alpha\beta}$. We will solve $\gamma^{\alpha\beta}$ using $P_u^\alpha$ and $\tau_\alpha$ as follows. The solution of this equation can be written as
 \cite{Harmark:2017rpg}
\begin{equation}
\sqrt{-\gamma}\gamma^{\alpha\beta}=e(-v^\alpha v^\beta+e^\alpha e^\beta) \ ,
\end{equation}
where
\begin{equation}
e_\alpha=\frac{\epsilon_{\alpha\beta}P^\beta_u}{T} \ , \quad
v^\alpha=-\frac{P^\alpha_u}{P_u^\gamma \tau_\gamma} \ , \quad
e^\alpha=-T \frac{\epsilon^{\alpha\beta}\tau_\beta}{P_u^\gamma \tau_\gamma} \ ,
\end{equation}
and where
\begin{equation}
e=
\det\left(\begin{array}{cc}
\tau_\tau & e_\tau \\
\tau_\sigma & e_\sigma \\ \end{array}\right)=\tau_\tau e_\sigma-
\tau_\sigma e_\tau=\epsilon^{\alpha\beta}\tau_\alpha e_\beta \ ,
\end{equation}
where we also  defined $\epsilon^{\tau\sigma}=-\epsilon_{\tau\sigma}=1$ so that
$\epsilon^{\alpha\gamma}
\epsilon_{\gamma\beta}=\delta^\alpha_\beta$.
Using this result we can write the Lagrangian density $\hat{\mL}$ as
\begin{equation}
\hat{\mL}=-\frac{T}{2}\sqrt{-\gamma}\gamma^{\alpha\beta}\hh_{\alpha\beta}=
-\frac{T}{2}e(-v^\alpha v^\beta+e^\alpha e^\beta)\hh_{\alpha\beta} \ ,
\end{equation}
where $\hh_{\mu\nu}=h_{\mu\nu}-m_\mu \tau_\nu-\tau_\mu m_\nu\ , \hh_{\alpha\beta}=\hh_{\mu\nu}\partial_\alpha x^\mu
\partial_\beta x^\nu$.

As the final step we solve the equation of motion (\ref{eqPu}).
This equation can be  solved locally by $e_\alpha=\partial_\alpha \eta$ and we substitute this result into the action $S=\int d^2\sigma \hat{\mL}$. Then we obtain following Lagrangian $\hat{\mL}$ in the form
%
%
\begin{eqnarray}\label{mLObers}
\hat{\mL}&=&-T\epsilon^{\alpha\beta}\partial_\beta \eta m_\mu \partial_\alpha x^\mu+
\nonumber \\
&+&\frac{T}{2}\frac{\epsilon^{\alpha\alpha'}\epsilon^{\beta\beta'}
    (\partial_{\alpha'}\eta\partial_{\beta'}\eta-\tau_\mu \partial_{\alpha'}x^\mu \tau_\nu\partial_{\beta'}x^\nu)}{\epsilon^{\gamma\gamma'}\tau_\mu\partial_\gamma x^\mu \partial_{\gamma'}\eta}h_{\mu\nu}\partial_\alpha x^\mu\partial_\beta x^\nu \ .
\end{eqnarray}
This Lagrangian density is the starting point for the canonical analysis.
\section{Canonical Analysis}\label{third}
In this section we perform canonical analysis of the Lagrangian density
(\ref{mLObers}). First of all we derive  following conjugate momenta
\begin{eqnarray}\label{peta}
p_\eta&=&\frac{\partial \hat{\mL}}{\partial (\partial_0 \eta)}=
Tm_\sigma -
T\frac{\epsilon^{\beta\beta'}
    \partial_{\beta'}\eta }{\epsilon^{\gamma\gamma'}
\tau_\gamma \partial_{\gamma'}\eta}h_{\sigma\beta}+
\frac{T}{2}\frac{\epsilon^{\alpha\alpha'}\epsilon^{\beta\beta'}
    (\partial_{\alpha'}\eta\partial_{\beta'}\eta-\tau_{\alpha'} \tau_{\beta'})}{(\epsilon^{\gamma\gamma'}\tau_\gamma  \partial_{\gamma'}\eta)^2}h_{\alpha\beta} \tau_\sigma \nonumber \\
\end{eqnarray}
and
\begin{eqnarray}\label{pmu}
p_\mu&=&\frac{\partial \hat{\mL}}{\partial (\partial_0 x^\mu)}=
-T\partial_\sigma \eta m_\mu
+T\frac{\epsilon^{\beta\beta'}
    (\partial_{\sigma}\eta\partial_{\beta'}\eta-\tau_\sigma \tau_{\beta'})}{\epsilon^{\gamma\gamma'}\tau_\gamma  \partial_{\gamma'}\eta}h_{\mu\nu}\partial_\beta x^\nu
\nonumber \\
&+&T\frac{\epsilon^{\beta\beta'}\tau_\mu \tau_{\beta'}
}{\epsilon^{\gamma\gamma'}\tau_\gamma \partial_{\gamma'}
    \eta}h_{\sigma \beta}-
\frac{T}{2}\frac{\epsilon^{\alpha\alpha'}\epsilon^{\beta\beta'}
    (\partial_{\alpha'}\eta\partial_{\beta'}\eta-\tau_{\alpha'} \tau_{\beta'})}{(\epsilon^{\gamma\gamma'}\tau_\gamma  \partial_{\gamma'}\eta)^2}h_{\alpha \beta}\tau_\mu \partial_\sigma \eta \ .
\nonumber \\
\end{eqnarray}
It is easy to see that   (\ref{peta}) and (\ref{pmu}) imply following primary constraint
\begin{eqnarray}\label{mHsigma}
\mH_\sigma=p_\eta \partial_\sigma \eta+p_\mu\partial_\sigma x^\mu\approx 0
\end{eqnarray}
while we find that the bare Hamiltonian is zero
\begin{eqnarray}
H_{bare}=p_\eta\partial_0 \eta+p_\mu \partial_0 x^\mu-\hat{\mL}=0
\end{eqnarray}
with agreement with the fact that the action $S=\int d^2\sigma \hat{\mL}$ is invariant under two dimensional diffeomorphism.
 On the other hand we are still missing
Hamiltonian constraint. In order to find it we have to introduce following
objects
 $e^\mu_{ \ a}$ and $v^\mu$ that are  defined as
\begin{eqnarray}
& &e_\mu^{ \ a}e^\mu_{ \ b}=\delta^a_b \ , \quad  e_\mu^{ \ a}e^\nu_{ \ a}=
\delta_\mu^\nu+\tau_\mu v^\nu \ , \quad  v^\mu\tau_\mu=-1 \ , \nonumber \\
& & v^\mu e_\mu^{ \ a}=0 \ , \quad  \tau_\mu e^\mu_{ \ a}=0 \ . \nonumber \\
\end{eqnarray}
The special and temporal vierbeins define special and temporal metrics as follows
\begin{eqnarray}
& &\tau_{\mu\nu}=\tau_\mu \tau_\nu \ , \quad \tau^{\mu\nu}=v^\mu v^\nu \ ,
\nonumber \\
& & h_{\mu\nu}=e_\mu^{ \ a}e_\nu^{ \ b}\delta_{ab} \ , \quad  h^{\mu\nu}=
e^\mu_{ \ a}e^\nu_{ \ b}\delta^{ab} \ . \nonumber \\
\end{eqnarray}
To proceed further we observe that we can write
\begin{eqnarray}
& &\frac{T^2}{
(\epsilon^{\gamma\gamma'}
\tau_\gamma\partial_{\gamma'}\eta)^2}\epsilon^{\beta\beta'}
    (\partial_{\sigma}\eta\partial_{\beta'}\eta-\tau_{\sigma}
    \tau_{\beta'})h_{\beta\alpha}
\epsilon^{\alpha\beta''}
    (\partial_{\sigma}\eta\partial_{\beta''}\eta-\tau_{\sigma}
    \tau_{\beta''})=\nonumber \\
& &=\frac{T^2}{(\epsilon^{\gamma\gamma'}
    \tau_\gamma \partial_{\gamma'}\eta)^2}
(\partial_\sigma \eta\partial_\sigma\eta-
\tau_\sigma \tau_\sigma)
\epsilon^{\alpha\alpha'}\epsilon^{\beta\beta'}
(\partial_{\alpha'} \eta\partial_{\beta'}\eta-
\tau_{\alpha'} \tau_{\beta'})
h_{\alpha\beta}
+T^2h_{\sigma\sigma} \
\nonumber \\
\end{eqnarray}
and also
\begin{eqnarray}
& &(p_\mu+T\partial_\sigma \eta m_\mu)v^\mu\partial_\sigma \eta
-(p_\eta-Tm_\mu\partial_\sigma x^\mu)\tau_\sigma=\nonumber \\
&=&\frac{T}{2(\epsilon^{\gamma\gamma'}\tau_\gamma \partial_{\gamma'}\eta)^2}
\epsilon^{\alpha\alpha'}\epsilon^{\beta\beta'}
(\partial_{\alpha'}\eta\partial_{\beta'}\eta-\tau_{\alpha'}
\tau_{\beta'})
h_{\alpha\beta}(\partial_\sigma \eta\partial_\sigma \eta-
\tau_\sigma\tau_\sigma)   +Th_{\sigma\sigma} \ .
\nonumber \\
\end{eqnarray}
If we combine these two relations together we obtain
 following primary Hamiltonian  constraint
\begin{eqnarray}
\mH_\tau&\equiv&
(p_\mu+T\partial_\sigma \eta m_\mu)h^{\mu\nu}
(p_\nu+T\partial_\sigma \eta m_\nu)-\nonumber \\
&-&
2T(p_\mu+T\partial_\sigma \eta m_\mu)v^\mu\partial_\sigma \eta
+2T(p_\eta-Tm_\mu\partial_\sigma x^\mu)\tau_\sigma+T^2 h_{\sigma\sigma}\approx 0 \ .
\nonumber \\
\end{eqnarray}
It is instructive to rewrite this constraint into the  form
\begin{eqnarray}\label{mHtauin}
\mH_\tau
=p_\mu h^{\mu\nu}p_\nu-2Tp_\mu\hv^\mu\partial_\sigma \eta+T^2\hh_{\mu\nu}\partial_\sigma x^\mu\partial_\sigma x^\nu+2Tp_\eta \tau_\mu
\partial_\sigma x^\mu+
2T^2\partial_\sigma \eta \Phi\partial_\sigma \eta \ , \nonumber \\
\end{eqnarray}
where
\begin{eqnarray}
\hh_{\mu\nu}=h_{\mu\nu}-m_\mu \tau_\nu-m_\nu\tau_\mu \ , \quad
\Phi=-m_\mu v^\mu+\frac{1}{2}m_\mu h^{\mu\nu}m_\nu \ , \quad
\hv^\mu=v^\mu-h^{\mu\nu}m_\nu \ .
\nonumber \\
\end{eqnarray}
These objects are  invariant under local Galilean transformations whose non-zero
transformation rules are
\begin{equation}
\delta e_\mu^{ \ a}=\tau_\mu \lambda^a \ ,
\quad
\delta v^\mu=e^\mu_{ \ a}\lambda^a \ ,
\quad
\delta m_\mu=e_\mu^{ \ a}\lambda_a \ ,
\end{equation}
where $\lambda_a$ is parameter of local Galilean transformations.

In summary, we have found that the Hamiltonian of non-relativistic string
is the sum of two primary constraints (\ref{mHsigma}) and
 (\ref{mHtauin}). In the next subsection we will analyze Poisson algebra of
these constraints.
\subsection{Algebra of Constraints}
As usual we have to determine an algebra of constraints. We define
smeared form of these constraints as
\begin{equation}
\bT_T(N)=\int d\sigma N\mH_\tau \ , \quad \bT_S(N^\sigma)=\int
d\sigma N^\sigma \mH_\sigma
\end{equation}
so that we have
\begin{eqnarray}\label{pbbTSbTS}
\pb{\bT_S(N^\sigma),\bT_S(M^\sigma)}=
\bT_S(N^\sigma \partial_\sigma M^\sigma-M^\sigma\partial_\sigma N^\sigma) \ .
\nonumber \\
\end{eqnarray}
To proceed further we calculate
\begin{eqnarray}\label{pbbTSmHtau}
\pb{\bT_S(N^\sigma),\mH_\tau}=-2\partial_\sigma N^\sigma \mH_\tau-
N^\sigma \partial_\sigma \mH_\tau \ , \nonumber \\
\end{eqnarray}
using
\begin{eqnarray}
& &\pb{\bT_S(N^\sigma),p_\mu}=-\partial_\sigma (N^\sigma p_\mu) \ , \quad
\pb{\bT_S(N^\sigma),x^\mu}=-N^\sigma \partial_\sigma x^\mu  \ ,
\nonumber \\
& &\pb{\bT_S(N^\sigma),p_\eta}=-\partial_\sigma (N^\sigma p_\eta) \ , \quad
\pb{\bT_S(N^\sigma),\eta}=-N^\sigma\partial_\sigma \eta \ .
\nonumber \\
\end{eqnarray}
Then  (\ref{pbbTSmHtau}) can be equivalently written as
\begin{equation}\label{pbbTSbTT}
\pb{\bT_S(N^\sigma),\bT_T(M)}=\bT_T(N^\sigma\partial_\sigma M-M\partial_\sigma N) \ .
\end{equation}
Finally we calculate Poisson bracket between smeared form of Hamiltonian constraints and we obtain
\begin{eqnarray}\label{pbbttth}
& &\pb{\bT_T(N),\bT_T(M)}=\int d\sigma (N\partial_\sigma M-M\partial_\sigma N)
4T^2 p_\nu h^{\nu\mu}\hh_{\mu\rho}\partial_\sigma x^\rho+\nonumber \\
&+&\int d\sigma (N\partial_\sigma M-M\partial_\sigma N)4Tp_\nu h^{\nu\mu}\tau_\mu p_\eta-\nonumber \\
&-&\int d\sigma (N\partial_\sigma M-M\partial_\sigma N)4T^3\hv^\mu\hh_{\mu\nu}
\partial_\sigma x^\nu\partial_\sigma \eta -\nonumber \\
&-&\int d\sigma (N\partial_\sigma M-M\partial_\sigma N)4T^2\hv^\mu \tau_\mu p_\eta \partial_\sigma \eta-\nonumber \\
&-&\int d\sigma (N\partial_\sigma M-M\partial_\sigma N)4T^2p_\mu \hv^\mu
\tau_\nu\partial_\sigma x^\nu+\nonumber \\
&+&\int d\sigma (N\partial_\sigma M-M\partial_\sigma N)8T^3 \tau_\mu \partial_\sigma x^\mu \Phi\partial_\sigma \eta \ . \nonumber \\
\end{eqnarray}
To proceed further we use the fact that
\begin{eqnarray}
& &h^{\mu\nu}\tau_\nu=0 \ , \quad  h^{\nu\mu}h_{\mu\rho}=\delta^\nu_\rho+v^\nu\tau_\rho \ ,
\nonumber \\
& & h^{\nu\mu}\hh_{\mu\rho}=\delta^\nu_\rho+\hv^\nu\tau_\rho \ , \quad
\hv^\mu\hh_{\mu\nu}=2\Phi\tau_\nu \ .\nonumber \\
\end{eqnarray}
Then if we combine these results together in (\ref{pbbttth}) we obtain desired result
\begin{equation}\label{pbbTTbTT}
\pb{\bT_T(N),\bT_T(M)}=
4T^2\bT_S(N\partial_\sigma M-M\partial_\sigma N) \ .
\end{equation}
We see that the Poisson brackets (\ref{pbbTSbTS}),(\ref{pbbTSbTT}) and
(\ref{pbbTTbTT}) close on the constraint surface $\mH_\sigma\approx 0 \ ,
\mH_\tau\approx 0$
 and hence
they are the  first class constraints.
\subsection{Fixing  Gauge}
We have seen that the non-relativistic string Hamiltonian is the sum of
two first class constraints. The natural way how to deal with such a theory is to gauge fix these constraints. For example, we can introduce  following gauge fixing functions
\begin{eqnarray}
\mG_\tau\equiv \sqrt{T}x^0-\tau \approx 0 \ , \quad
\mG_\sigma \equiv \sqrt{T}\eta-\sigma \approx 0 \ .
\nonumber \\
\end{eqnarray}
To see this that they are suitable gauge fixing functions we calculate
following  Poisson brackets
\begin{eqnarray}
& &\pb{\mG_\tau(\sigma),\mH_\tau(\sigma')}\approx
(2\sqrt{T}h^{0\nu}p_\nu-2T\hv^{0})\delta(\sigma-\sigma') \ , \quad
\pb{\mG_\tau(\sigma),\mH_\sigma(\sigma')}\approx 0 \ , \nonumber \\
& &\pb{\mG_\sigma(\sigma),\mH_\tau(\sigma')}=2T^{3/2}\tau_\mu\partial_\sigma x^\mu\delta(\sigma-\sigma') \ , \quad
\pb{\mG_\sigma(\sigma),\mH_\sigma(\sigma')}
\approx \delta(\sigma-\sigma') \ . \nonumber \\
\end{eqnarray}
Since these Poisson brackets do not vanish on the constraint surface
$\mH_\tau\approx 0 \ ,\mH_\sigma\approx 0 \ , \mG_\tau\approx 0 \ , \mG_\sigma\approx 0$ we see that these gauge fixing functions together with
$\mH_\tau\approx 0 \ ,  \mH_\sigma\approx 0$ are the second class constraints that vanish strongly. Then the Hamiltonian on the reduced phase space  follows from the action
\begin{eqnarray}\label{gaugefixedact}
S&=&\int d^2\sigma (p_\mu\partial_0 x^\mu+p_\eta \partial_0 \eta-
\lambda^\tau \mH_\tau-\lambda^\sigma \mH_\sigma)=
\nonumber \\
&=&\int d^2\sigma (p_i \partial_\tau x^i+\frac{1}{\sqrt{T}}p_0) \ ,
\end{eqnarray}
where we used the fact that $\mH_\tau=0,\mH_\sigma=0$ and we also  used
$\mG_\tau=0$ to express $x^0$ as $x^0=\frac{1}{\sqrt{T}}\tau$. Then we see
from (\ref{gaugefixedact}) that it is natural to identify the gauge fixed Hamiltonian density as
\begin{equation}
\mH_{fixed}=-\frac{1}{\sqrt{T}}p_0 \ ,
\end{equation}
where  $p_0$ can be determined from $\mH_\tau=0$, at least in principle, while from $\mH_\sigma=0$ we obtain $p_\eta$ as
\begin{equation}
p_\eta=-\sqrt{T}p_i\partial_\sigma x^i
\ .
\end{equation}
Let us consider for example a flat Newton-Cartan background when $m_\mu=0 \ , \tau_\mu=\delta^0_\mu, h_{\mu\nu}=\delta_{ij}\delta^i_\mu \delta^j_\nu$. Then clearly $h^{\mu\nu}=\delta^\mu_i\delta^\nu_j\delta^{ij}, \hv^\mu=-\delta^\mu_0$
so that
\begin{eqnarray}
\mH_\tau
=p_i \delta^{ij}p_j+2\sqrt{T}p_0
+T^2\delta_{ij}\partial_\sigma x^i\partial_\sigma x^j=0
\nonumber \\
\end{eqnarray}
that can be easily solved for $p_0$. As a result we obtain
 the Hamiltonian density on the reduced phase space in the form
\begin{equation}
\mH_{fixed}=\frac{1}{T}p_i \delta^{ij}p_j+T\delta_{ij}\partial_\sigma x^i
\partial_\sigma x^j \ .
\end{equation}

\subsection{Scaling Limit}
It was shown in  \cite{Harmark:2017rpg} that the second scaling
limit $T\rightarrow 0$ defines new interesting class of
non-relativistic theories. Let us now implement this idea in case of
the Hamiltonian formulation of this theory. From the form of the
constraint $\mH_\tau\approx 0$ it is clear that the naive limit
$T\rightarrow 0$ in the constraint $\mH_\tau\approx 0$
 leads to a trivial dynamics since $\mH_\tau\rightarrow p_\mu h^{\mu\nu}p_\nu\approx 0$ for $T\rightarrow 0$.  In order to resolve this problem
 we follow the analysis proposed in  \cite{Harmark:2017rpg}. Explicitly,  in order to  make the Hamiltonian constraint non-trivial we have to  rescale the coupling to $v^\mu$ too.  In more details, let us write $\tau_\mu$ as
\begin{equation}
\tau_\mu=N\partial_\mu F+\beta_\mu \ , \quad  v^\mu \beta_{\mu}=v^\mu h_{\mu\nu}=0 \ , \quad v^\mu \tau_\mu=-1 \ .
\end{equation}
Let us consider scaling limit
\begin{equation}
F=c^2\tF \ , \quad T=\frac{\tilde{T}}{c} \ , \quad
\eta=c\tilde{\eta} \ , p_\eta=\frac{1}{c}p_{\tilde{\eta}} \ , \quad c\rightarrow \infty \ .
\end{equation}
Further, since  $v^\mu\tau_\mu=-1$ that holds for all $c$ we should rescale $v^\mu$ as  $v^\mu=\frac{1}{c^2}\tilde{v}^\mu$.
With the help of this prescription we find that the Hamiltonian constraint
scales as
\begin{eqnarray}\label{mHtauscal}
\mH_\tau
\Rightarrow \tilde{\mH}_\tau=
(p_\mu+\tT\partial_\sigma\teta m_\mu)h^{\mu\nu}(p_\nu+\tT\partial_\sigma \teta m_\nu)+2\tT(p_\eta-\tT m_\mu\partial_\sigma x^\mu)\ttau_\sigma\approx 0 \ ,  \nonumber \\
\end{eqnarray}
where  $\ttau_{\mu}=N\partial_\mu F $.

As a check whether our approach is correct let us start with the scaled action found in \cite{Harmark:2017rpg}
\begin{equation}\label{Actlim}
S=-\tT\int d^2\sigma
\left(\epsilon^{\alpha\beta}m_{\alpha}
\partial_\beta \teta+\frac{\epsilon^{\alpha\alpha'}\epsilon^{\beta\beta'}
\ttau_{\alpha'}   \ttau_{\beta'}}
{2\epsilon^{\gamma\gamma'}\ttau_{\gamma}
\partial_{\gamma'}\teta}    h_{\alpha\beta}
    \right)
\end{equation}
and determine corresponding Hamiltonian. From  (\ref{Actlim})
we derive following conjugate momenta
\begin{eqnarray}
p_{\teta}&=&\frac{\partial \tilde{\mL}}{\partial \partial_0 \teta}=
\tT m_\sigma-\tT
\frac{\epsilon^{\alpha\alpha'}\epsilon^{\beta\beta'}
    \ttau_{\alpha'}  \ttau_{\beta'}}
{2(\epsilon^{\gamma\gamma'}\ttau_{\gamma}
    \partial_{\gamma'}\teta)^2} h_{\alpha\beta}\ttau_\sigma
\nonumber \\
p_\mu&=&\frac{\partial \tilde{\mL}}{\partial\partial_0 x^\mu}=-\tT m_\mu \partial_\sigma \teta
+\tT\frac{\epsilon^{\beta\beta'}
    \ttau_\mu \ttau_{\beta'}}
{\epsilon^{\gamma\gamma'}\ttau_{\gamma}
    \partial_{\gamma'}\teta}    h_{\sigma\beta}+
\nonumber \\
&+&\frac{\epsilon^{\alpha\alpha'}\epsilon^{\beta\beta'}
    \ttau_{\alpha'}   \ttau_{\beta'}}
{2(\epsilon^{\gamma\gamma'}\ttau_{\gamma}
    \partial_{\gamma'}\teta)^2} h_{\alpha\sigma}    \ttau_\mu\partial_\sigma\teta
-\tT
\frac{\epsilon^{\beta\beta'}
    \ttau_{\sigma}  \ttau_{\beta'}}
{\epsilon^{\gamma\gamma'}\ttau_{\gamma}
    \partial_{\gamma'}\teta}    h_{\mu\sigma}
\partial_\sigma x^\sigma     \ .
\nonumber \\
\end{eqnarray}
Then performing the same manipulation as in previous section we derive
following Hamiltonian constraint
\begin{equation}
\mH_\tau=
(p_\mu+\tT m_\mu\partial_\sigma \teta)h^{\mu\nu}
(p_\nu+\tT m_\nu\partial_\sigma \teta)+2\tT(p_{\teta}-\tT m_\mu
\partial_\sigma x^\mu)\ttau_\sigma\approx 0 \
\end{equation}
that coincides with (\ref{mHtauscal}).
\section{Alternative Derivation of Non-Relativistic String}\label{fourth}
In this section we perform an alternative derivation of the
non-relativistic string in the Newton-Cartan background. The
starting point of our construction is the  Hamiltonian for the
string  in the  background with null isometry. We begin with
Nambu-Goto form of the string action in general background
\begin{equation}\label{actionNG}
S=-T\int d^2\sigma \sqrt{-\det g_{\alpha\beta}}
\end{equation}
and find its Hamiltonian form.  From (\ref{actionNG})
we obtain conjugate momenta
\begin{equation}
p_M=-T G_{MN}\partial_\alpha x^N g^{\alpha\tau}\sqrt{-\det
    g}  \ .
\end{equation}
Using this relation it is easy to find two primary constraints
\begin{eqnarray}\label{mHtaugen}
\mH_\tau=
p_M G^{MN}p_N+T^2 G_{MN}\partial_\sigma x^M\partial_\sigma x^N\approx 0
\ ,  \quad
\mH_\sigma=p_M\partial_\sigma x^M \ .  \nonumber \\
\end{eqnarray}
As in section (\ref{second}) we now consider  background metric with null isometry
\begin{equation}
ds^2=G_{MN}dx^M dx^N=2\tau(du-m)+h_{\mu\nu}dx^\mu dx^\nu \ ,
\end{equation}
where
\begin{equation}
\tau=\tau_\mu dx^\mu \ , \quad  m=m_\mu dx^\mu \ ,
\end{equation}
and where $\det h_{\mu\nu}=0$.
It can be shown that the inverse metric $G^{MN}$ has the form
\begin{equation}
G^{uu}=2\Phi \ , \quad G^{u\mu}=-\hv^\mu \ , \quad
G^{\mu\nu}=h^{\mu\nu} \ .
\end{equation}
In this background the Hamiltonian and diffeomorphism constraints (\ref{mHtaugen}) have the form
\begin{eqnarray}
\mH_\tau&=&2p_u \Phi p_u-2p_u \hv^\mu p_\mu+p_\mu h^{\mu\nu}p_\nu+
2T^2\tau_\mu \partial_\sigma x^\mu \partial_\sigma u+
T^2\hh_{\mu\nu}\partial_\sigma x^\mu
\partial_\sigma x^\nu \ ,
\nonumber \\
\mH_\sigma&=& p_u\partial_\sigma u+p_\mu\partial_\sigma x^\mu \ . \nonumber \\
\end{eqnarray}
This is the Hamiltonian constraint for the string in the null background. Note that this background possesses an isometry
\begin{equation}
u\rightarrow u+\epsilon \ , \quad \epsilon=\mathrm{const} \ .
\end{equation}
Let us now perform canonical transformation from $u$ to $\eta$
\cite{Alvarez:1994wj,Alvarez:1994dn}
 when we
presume that the generating function has the form
\begin{equation}\label{genfuncT}
G=\frac{T}{2}\int d\sigma (u\partial_\sigma \eta-\partial_\sigma u\eta) \ .
\end{equation}
Let us denote the momentum conjugate to $\eta$ as $p_\eta$. Then from
the definition of the canonical transformation we derive
following relation between  momenta $p_u, p_\eta$ and spatial derivatives
of $u$ and $\eta$:
\begin{eqnarray}
p_\eta&=&-\frac{\delta G}{\delta \eta}=T\partial_\sigma u \ ,
\quad
p_u= \frac{\delta G}{\delta u}=T\partial_\sigma \eta \ .
\nonumber \\
\end{eqnarray}
Now we obtain canonically dual Hamiltonian when we replace
$\partial_\sigma u$ with $\frac{1}{T}p_\eta$ and
$p_u$ with $T\partial_\sigma \eta$ in $\mH_\tau$ and $
\mH_\sigma$ given above and we obtain
\begin{eqnarray}\label{mHconst}
& &\mH_\sigma=p_\eta\partial_\sigma \eta+p_\mu\partial_\sigma x^\mu\approx 0 \ , \nonumber \\
& &\mH_\tau=p_\mu h^{\mu\nu}p_\nu-2Tp_\mu \hv^\mu\partial_\sigma \eta
+2\tau_\mu\partial_\sigma x^\mu p_\eta+T^2 \hh_{\mu\nu}\partial_\sigma x^\mu
\partial_\sigma x^\nu+2T^2\partial_\sigma \eta \Phi
\partial_\sigma \eta \ .
\nonumber \\
\end{eqnarray}
which precisely coincide with the Hamiltonian constraint
(\ref{mHtauin}). It is important to stress that the canonical transformation
defined by the generating function
(\ref{genfuncT})  can be interpreted as T-duality transformation along
$u-$direction even if this interpretation is slightly formal due to the fact
that the $u-$direction is null and hence non-compact.

Finally we check our result by derivation of the  Lagrangian density from the
Hamiltonian $H=\int d\sigma (\lambda^\tau \mH_\tau+\lambda^\sigma \mH_\sigma)$,
where the constraints $\mH_\tau\approx 0 \ , \mH_\sigma\approx 0$ are given
in (\ref{mHconst}). Using this Hamiltonian we easily find
\begin{eqnarray}
\partial_0 x^\mu&=&\pb{x^\mu,H}=2\lambda^\tau h^{\mu\nu}p_\nu-2\lambda^\tau T\hv^\mu
\partial_\sigma \eta+\lambda^\sigma \partial_\sigma x^\mu \ , \nonumber \\
\partial_0 \eta&=&\pb{\eta,H}=2\lambda^\tau\tau_\mu \partial_\sigma x^\mu +\lambda^\sigma \partial_\sigma \eta\nonumber \\
\end{eqnarray}
and we obtain following Lagrangian density
\begin{eqnarray}
\mL&=&p_\mu \partial_0 x^\mu+p_\eta\partial_0 \eta-\lambda^\tau \mH_\tau-
\lambda^\sigma \mH_\sigma=\nonumber \\
&=&\lambda^\tau p_\mu h^{\mu\nu}p_\nu-\lambda^\tau T^2\hh_{\mu\nu}\partial_\sigma x^\mu
\partial_\sigma x^\nu-2\lambda^\tau T^2\partial_\sigma \eta \Phi\partial_\sigma \eta  \ . \nonumber  \\
\end{eqnarray}
To proceed further we introduce $\he_\mu^{ \ a}$ as
\begin{equation}
\he_\mu^{ \ a}=e_\mu^{ \ a}-e^\nu_{ \ b}\delta^{ba}m_\nu \tau_\mu
\end{equation}
that obeys
\begin{equation}
\he_\mu^{ \ a}h^{\mu\nu}=\delta^{ab}e^\nu_{ \ b}  \ .
\end{equation}
Using this relation we easily find
\begin{eqnarray}
p_\mu h^{\mu\nu}p_\nu=
\frac{1}{4(\lambda^\tau)^2}
(\partial_\tau x^\mu+2T\hv^\mu\partial_\sigma \eta-\lambda^\sigma \partial_\sigma x^\mu)\he_\mu^{ \ a}\delta_{ab}
\he_\nu^{ \ b}
(\partial_\tau x^\nu+2T\hv^\nu\partial_\sigma \eta-\lambda^\sigma \partial_\sigma x^\nu) \nonumber \\
\end{eqnarray}
and hence we obtain Lagrangian density in the form
\begin{eqnarray}
\mL&=&\frac{1}{4\lambda^\tau}
(\partial_\tau x^\mu+2T\hv^\mu\partial_\sigma \eta-\lambda^\sigma \partial_\sigma x^\mu)\he_\mu^{ \ a}\delta_{ab}
\he_\nu^{ \ b}
(\partial_\tau x^\nu+2T\hv^\nu\partial_\sigma \eta-\lambda^\sigma \partial_\sigma x^\nu)
-\nonumber \\
&-&\lambda^\tau T^2\hh_{\mu\nu}\partial_\sigma x^\mu
\partial_\sigma x^\nu-2\lambda^\tau T^2\partial_\sigma \eta \Phi\partial_\sigma \eta \ .  \nonumber  \\
\end{eqnarray}
This Lagrangian density can be rewritten into an equivalent form
if we use the relation
\begin{eqnarray}
\he_\mu^{ \ a}\delta_{ab}\he_\nu^{ \ b}=
\hh_{\mu\nu}+2\tau_\mu \Phi\tau_\nu
\end{eqnarray}
so that the  Lagrangian density has the form
\begin{eqnarray}
\mL
&=&\frac{1}{4\lambda^\tau}(\hh_{\tau\tau}+2\tau_\tau \Phi\tau_\tau-2\lambda^\sigma \hh_{\tau\sigma}-4\lambda^\sigma
\tau_\tau \Phi\tau_\sigma+\nonumber \\
&+&(\lambda^\sigma)^2\hh_{\sigma\sigma}+
(\lambda^\sigma)^2 2\tau_\sigma \Phi\tau_\sigma)-
\lambda^\tau T^2\hh_{\sigma\sigma}-2\lambda^\tau T^2\partial_\sigma \eta
\Phi\partial_\sigma \eta \ . \nonumber \\
\end{eqnarray}
Finally we  eliminate
$\lambda^\tau, \lambda^\sigma$ from $\mL$. As was argued in
\cite{Kluson:2018uss} these multipliers cannot be eliminated by their
equations of motion. Instead we have to examine
 the equations of motion for $x^\mu$ and $\eta$. In fact, from the  equation of motion for $x^\mu$ we obtain
\begin{equation}\label{tautau}
\tau_\tau=2\lambda^\tau \partial_\sigma \eta+\lambda^\sigma\tau_\sigma
\end{equation}
while from the equation of motion for $\eta$ we obtain
\begin{equation}
\frac{1}{2\tau_\sigma}
(\partial_\tau \eta-\lambda^\sigma\partial_\sigma\eta)=\lambda^\tau \ .
\end{equation}
Inserting this result into (\ref{tautau})
 we obtain
\begin{equation}
\lambda^\sigma=\frac{\tau_{\tau\sigma}-\partial_\tau \eta\partial_\sigma\eta}{\tau_{\sigma\sigma}-\partial_\sigma
\eta\partial_\sigma\eta}
\end{equation}
and hence we find that $\lambda^\tau$ is equal to
\begin{equation}
\lambda^\tau
=-\frac{\epsilon^{\alpha\beta}\tau_\alpha\partial_\beta\eta}{2(\tau_\sigma\tau_\sigma-\partial_\sigma \eta\partial_\sigma\eta)} \ .
\end{equation}
With the help of these results it is easy to find the Lagrangian
density in the form
\begin{eqnarray}\label{mLfromH}
\mL
&=&-\frac{1}{2\epsilon^{\gamma\gamma'}\tau_\gamma\partial_{\gamma'}\eta}
(\tau_\alpha\tau_\beta-\partial_\alpha\eta\partial_\beta\eta)
\epsilon^{\alpha\alpha'}\epsilon^{\beta\beta'}\hh_{\alpha'\beta'}
\nonumber \\
&-&\frac{1}{\epsilon^{\gamma\gamma'}\tau_{\gamma}\partial_{\gamma'}\eta}
(\tau_\alpha\tau_\beta-\partial_\alpha\eta\partial_\beta\eta)
\epsilon^{\alpha\alpha'}\epsilon^{\beta\beta'}
\tau_{\alpha'}\tau_{\beta'}\Phi-T^2\epsilon^{\alpha\beta}\tau_\alpha\partial_\beta\eta
\Phi\ .  \nonumber  \\
\end{eqnarray}
At first sight we should say
 that this Lagrangian density is different from the
one derived in \cite{Harmark:2017rpg}. However when we perform
closer examination we find that two expressions on the second line in
(\ref{mLfromH}) cancel each other and it precisely reduces into
(\ref{mLObers}).
We mean that this is really nice consistency check.
\section{Conclusion}\label{fifth}
Let us outline our results and suggest possible extension of this work.
We analyzed non-relativistic string theory on Newton-Cartan background
which was introduced in \cite{Harmark:2017rpg}.  We found its
Hamiltonian form and calculated an algebra of constraints. We also discussed
its gauge fixed form. We also shown an alternative way of the
derivation of this theory with the help of T-duality along null
direction. We mean that this is very interesting result that clearly allows
natural extension of this work when we analyze Green-Schwarz superstring
in the background with null isometry and perform T-duality along
this direction. It would be also nice to analyze the
action \cite{Harmark:2017rpg} in some specific background and
try to find solutions of corresponding equations of motion.
We hope to return to these problems in future.

\acknowledgments{This  work  was
    supported by the Grant Agency of the Czech Republic under the grant
    P201/12/G028. }


\end{document}